\documentclass[aps,floatfix]{revtex4}
\usepackage{graphics,epsfig}
\usepackage{graphicx}
\usepackage{amssymb,latexsym}
\usepackage{amsmath, amsthm}
\usepackage{amscd}
\begin{document}

\title{Confinement and stability of dynamical system in presence of scalar fields and perturbation in the bulk}

       % if too long for running head

\author{Pinaki Bhattacharya}
\affiliation{Gopal Nagar High School, Singur 712409, India
\\ Jadavpur University, Kolkata, India}
\author{Sarbari Guha}
\affiliation{Department of Physics, St. Xavier's College (Autonomous), 30 Mother Teresa Sarani, Kolkata 700 016,India}

\begin{abstract}
In this paper we have considered a five-dimensional warped product spacetime with spacelike extra dimension and with a scalar field source in the bulk. We have studied the dynamics of the scalar field under different types of potential in an effort to explain the confinement of particles in the five-dimensional spacetime. The behavior of the system is determined from the nature of damping force on the system. We have also examined the nature of the effective potential under different circumstances. Lastly we have studied the system to determine whether or not the system attains asymptotically stable condition for both unperturbed and perturbed conditions. The analysis throws significant light on the nature confinement of particles and the stability of the dynamical system under these conditions.
\end{abstract}

\maketitle
\section{Introduction}
\bigskip
Recent higher dimensional theories of gravity \cite{ru1,ru2} have put forward the idea that extra spatial dimensions possibly exist, but these are hidden from our observations due to the effect of warping instead of the traditional Kaluza-Klein compactification. This idea opened up new horizons in the field of particle physics and cosmology. The extra dimensions could be much larger than the Planck length, with ordinary matter of the observed four-dimensional (4D) universe being confined to a 4D sub-manifold with three spatial dimensions \cite{rs1}. The 4D hypersurface (called the `brane'), is warped by a function of the extra dimension (called the ‘warp factor’), to be embedded in the higher dimensional background. The simplest example of such warped braneworld models is the thin brane model of Randall and Sundrum (RS) \cite{rs2,rs3}. In this model, the brane is represented by a delta function term in the higher dimensional bulk, and the brane tension is responsible for the discontinuity in the first derivative of the metric at the junction of the two regions. Subsequently, other models with thin or thick branes have been constructed in five (or higher) dimensions \cite{rs4,rs5,rs6,rs7,rs8,rs9}. The thick brane models assume the existence of bulk fields other than pure gravity in the form of real scalars, yielding regular domain wall solutions having a finite thickness, thereby giving rise to well-behaved differentiable metrics. On account of 4D Poincar\'{e} invariance, these bulk fields acquire static nonzero configurations along the extra dimensions.

In the RS type II model, the zero mode perturbation of gravity is localized on the brane which is embedded in a non-compact five-dimensional (5D) anti-de Sitter ($AdS_5$) spacetime having mirror symmetry. As gravity leaks into the higher dimensional bulk, only a part of it is felt in the observed 4D universe. Confinement of gravity can be achieved in the thick brane versions of the RSII model \cite{rs9a,rs9b}. However, if the geodesic motion of particles gets perturbed, then stable confinement of particles cannot be achieved even in the RSII model \cite{rs9c}. As gravitational force alone is not sufficient for providing particle confinement in presence of perturbed motion, some additional mechanism is necessary to provide confinement in such a case.

In higher dimensional cosmology, scalar fields help us to explain different phenomena of gravitation. Scalar fields represent spin-0 particles in field theories, for example the Higgs field in the standard model of particle physics. In cosmology, the present acceleration of the universe is explained in terms of a so-called `dark energy'. Generally, a scalar field (e.g. quintessence) is considered as the possible source of dark energy. Other than the dark energy problem, different phases of the inflation \cite{Linde}, embedding (as in the braneworld models), can also be explained in terms of scalar fields \cite{rs9i,rs9j,rs9k,rs10}. Scalar fields can act as an `effective' cosmological constant driving an inflationary period of the universe. To produce a proper phase of inflation, the matter content of the universe must be dominated by a fluid with negative pressure. At very high energies, as in the initial phase of cosmological evolution, the correct description of matter is given by field theory in terms of a scalar field. In the simplest models of inflation, the energy density responsible for inflation arises from the expectation value of a scalar field rolling down a potential. If the potential of this scalar field is sufficiently flat so that the field moves slowly, then the corresponding pressure is negative. So it is believed that inflation is driven by one (or more) scalar field(s), called the “inflaton”. However, the shape of its potential is not known except that it must be sufficiently flat. This led to the consideration of several types of scalar field models, each having its own merits and demerits. Scalar fields exhibit different cosmological behaviours due to different types of self interactions, or potentials.

In most braneworld models, depending on the nature of the bulk, either the scalar field is assumed to be a function of the extra dimension as well as of the coordinates on the brane \cite{rs11} or is simply dependent on the extra dimensional coordinate. On a large scale, the Universe appears to be homogeneous. Therefore a cosmological scalar field is assumed to be space-independent, and only varies with time. In some cases, scalar fields have been used to explain the localization of fermions through Yukawa-type interactions, as in the Rubakov-Shaposhnikov model \cite{rs12}. A non-quantum description of stable confinement of particles in the brane due to a direct interaction between a bulk scalar field and the particles, was provided by Dahia et. al. \cite{rs12a}. They assumed the bulk scalar field to depend only on the extra-dimensional coordinate. These scalar fields are known as kinklike scalar fields \cite{rs12b}. In a field theoretic description of thick branes, such scalar fields are suitable for localizing bulk fermions and other scalar or vector fields near the 4D hypersurface \cite{DGV}.

In this work we have studied the behaviour of a scalar field in a 5D bulk with spacelike extra dimension, in order to derive an idea about the nature of particle confinement in such a 5D spacetime. We have assumed the scalar field to depend on the extra-dimensional coordinate only. On the other hand, the 5D bulk is a warped product space with the warping function dependent on the extra-dimensional coordinate. We divide our study into three parts. First, we have examined the nature of the scalar field and the corresponding `damping force' on the 4D brane. During this investigation we have showed how the warping function and the scalar field potential determine the nature of the `damping force', which helps us to study the nature of confinement of particles. We have assumed a massive scalar field so that we can consider different types of potential and determine the effect of perturbation on the field potential. In the second part we have studied the effective potential near the brane for a Yukawa type interaction between the test particle and the scalar field \cite{rs12a}. In the last section we have examined the stability of the dynamical system. We have discussed the effect of perturbation on the scalar field for an unwarped situation and showed how the nature of this perturbation gives rise to a limit cycle i.e an asymptotically stable situation. We have studied the system for both quadratic scalar field potential and the Higgs potential.

\section{Mathematical Preliminaries}
Let us consider a 5D line element represented by
\begin{equation}\label{01}
  dS^2 = e^{2f(y)}\left(dt^{2} -R^{2}(t)(dr^{2} + r^{2}d\theta^{2} + r^2sin^{2}\theta d\phi^{2})\right) - dy^{2},
\end{equation}
where the warping function is $f(y)$ and $y$ is a spacelike extra dimension. According to our assumption, the 5D manifold is foliated by a family of hypersurfaces defined by $y=\textrm{constant}$. So the geometry of the hypersurface at $y = y_{0}$ can be determined from the equation
\begin{equation}\label{02}
ds^2 =e^{2f({y_0})}\left(dt^2 -R^{2}(t)(dr^2 + r^2d\theta^2 + r^2sin^{2}\theta d\phi^2)\right).
\end{equation}
In the following expressions, a prime stands for derivative with respect to $y$.

A warped product space \cite{rs13,rs14}, can be defined as follows: Let us consider two manifolds (Riemannian or semi-Riemannian) $(M^{m}, h)$ and $(M^{n}, \bar{h})$ of dimensions $m$ and $n$, with metrics $h$ and $\bar{h}$ respectively. With the help of a smooth function $f : M^{n} \rightarrow \Re$ (henceforth referred to as the ’warping function’), we can build a new Riemannian (or semi-Riemannian) manifold $(M,g)$ by setting $M = M^{m} \times M^{n}$, which is defined by the metric $g = e^{2f}h \oplus \bar{h}$. Here (M, g) is called a ’warped product manifold’. For dim $M=4$, with signature $\pm 2$, (M, g) corresponds to a spacetime, called a ’warped product spacetime’, which in our case is a (3+1)-dimensional Lorentzian spacetime with signature (+,-,-,-).

\section{Scalar field and the damping force}
The nature of the scalar field is determined from the corresponding field equations. Let us consider the simplest case of
a one-component real scalar field $\phi$ minimally coupled to gravity in a warped extra dimension, where we neglect the back reaction produced by the scalar field on the warping, as considered by Toharia et al in \cite{to2}. The 5D scalar field equations are thus given by
\begin{eqnarray}
% \nonumber to remove numbering (before each equatio
\frac{1}{\sqrt{g}}\partial_{A}(\sqrt{g}g^{AB}\partial_{B}\phi)= -\frac{dV(\phi)}{d\phi},
\end{eqnarray}
where $A,B, ... = 0, 1, 2, 3, 4$. The vacuum expectation value (VEV) of the scalar field need not be constant, so that it can have a nonzero value along the extra dimension, given by $\langle \phi(x,y) \rangle = \phi_c(y)$, where $x$ is an element of the 4D hypersurface. It has been demonstrated \cite{ru1} that such a $y$-dependence of the VEV of a scalar field may lead to the natural localization of fermions living on the higher-dimensional bulk to the lower-dimensional hypersurface in the thick brane scenario.

The equation of motion for the scalar field now reduces to the form
\begin{eqnarray}
% \nonumber to remove numbering (before each equatio
\frac{1}{\sqrt{g}}\partial_{y}(\sqrt{g}g^{yy}\partial_{y}\phi)= -\frac{dV(\phi)}{d\phi}.
\end{eqnarray}
The evolution of the static nonzero configurations of the scalar field in the background spacetime considered by us is therefore given by the equation
\begin{equation}\label{04}
% \nonumber to remove numbering (before each equatio
\phi''(y)+ 4f'(y)\phi'(y) +\frac{dV(\phi)}{d\phi}= 0.
\end{equation}
This equation looks like the equation of standard damped harmonic motion. However, the difference is that here we are dealing with an equation for space evolution along the extra dimension at a given instant of time, instead of an actual time evolution. Thus the second term of \eqref{04} is analogous to a dissipative term, with the quantity $f'(y)$ being analogous to a `coefficient of friction'. The presence of the `dissipative' term makes this system `nonconservative' in nature. We know that during cosmological phase transition the scalar field is required to interact with other fields, to facilitate the transfer of energy from potential energy into radiation \cite{sa1,sa2}. A similar study can be done for a 5D bulk considering the type of scalar field described in \cite{to1a,to1b,to2}. Here we have considered a RSII type scenario with a thick brane. Assuming different types of scalar field potentials we have studied the nature of the `dissipative' force and its effect on the dynamical system, in order to get an idea of confinement in all these cases. The sample field potentials which we have considered are the following: (i) quadratic potential $V(\phi) = \frac{1}{2}m^{2}\phi^{2}$, where $m$ is the mass of the scalar field \cite{quadratic}, (ii) quartic potential $V(\phi) = \frac{1}{2}m^{2}\phi^{2} +\frac{\lambda}{4} \phi^{4}$ \cite{quartic}, where $\lambda$ is a coupling constant and (iii) $V(\phi) = \frac{C}{\phi^{2}}$ where C is a constant \cite{rs15}. For the quartic potential, it is known that the classical equations of motion of a scalar field admit a domain wall solution which is independent of the coordinates on the brane and depends only on the extra-dimensional coordinate \cite{ru1}. Its form coincides with (1 + 1)-dimensional kink solutions \cite{ru3} and provides a suitable mechanism for the trapping of ordinary matter. The study of this Mexican hat potential is a matter of great interest, especially in spontaneous symmetry breaking.

\subsection{The potential $V(\phi) = \frac{1}{2}m^{2}\phi^{2}$}
Let us consider the potential due to a massive scalar field. It is known that a lot of inflationary models are based on scalar fields because scalar fields play an important role in high energy physics. A massive scalar field helps us to describe chaotic inflation \cite{chao}. During chaotic inflation the scalar field is assumed not to be homogeneous. At the Planck time, the scalar field in some parts of the universe is large and in other parts of the universe the scalar field is low. So in those parts of the universe where the scalar field is large at the Planck time, inflationary expansion can take place. On other hand the places where the scalar field is low at the Planck time, will not experience inflationary expansion. So it is clear that under such a scenario, dissipation plays an important role in the stabilization of extra dimension. Keeping this in mind, here we shall study the nature of the dissipation in the bulk along the extra dimensional coordinate.

\begin{figure}[ht]
\begin{minipage}[b]{0.45\linewidth}
  \includegraphics[height=1.5in]{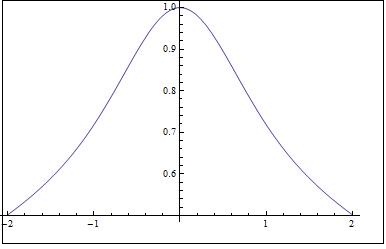}
  \caption{Figure describing the variation of the scalar field with respect to the extra dimensional coordinate when $V(\phi) = \frac{1}{2}m^{2}\phi^{2}$ and $f(y)=A\ln\cosh(y)$.}
  \label{Fig 1a}
  \end{minipage}
\hspace{0.5cm}
\begin{minipage}[b]{0.45\linewidth}
\includegraphics[height=1.5in]{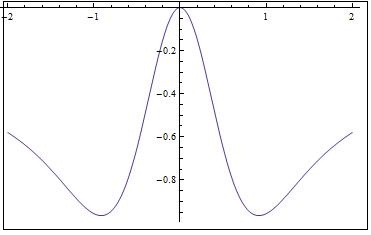}
\caption{\label{Fig 1b}Figure showing the variation of the dragging force with the extra dimensional coordinate for $V(\phi) = \frac{1}{2}m^{2}\phi^{2}$ and $f(y)=A\ln\cosh(y)$.}
\end{minipage}
\end{figure}

The scalar field equation in this circumstance becomes
 \begin{equation}\label{05}
% \nonumber to remove numbering (before each equatio
\phi''(y)+ 4f'(y)\phi'(y) + m^{2}\phi(y)= 0.
\end{equation}
Since we were unable to find the analytical solution for \eqref{05},  we solved the system in the same way as done in \cite{rs16} using the NDSolve routine from Mathematica, assuming the boundary condition $\phi(0)=1$ and $\phi'(0)=0$. FIG.~\ref{Fig 1a} and FIG.~\ref{Fig 1b} shows the nature of the scalar field and the nature of the corresponding damping force under a growing warping function $(f(y)=A\ln\cosh(y)$).
We find that the damping force is negative along the extra dimension but it increases rapidly near the brane. So the figure suggests that as we move towards the $y = 0$ hypersurface, the system will become more stable due to the effect of damping. However this mechanism may not provide efficient trapping of matter.

\begin{figure}[ht]
\includegraphics[height=1.5in]{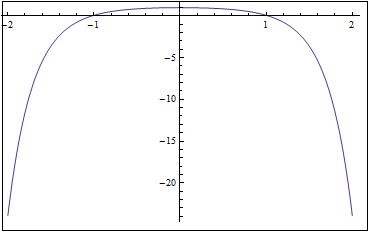}
\includegraphics[height=1.5in]{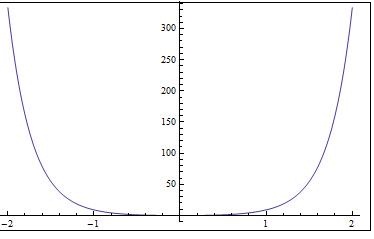}
\caption{The figure on the left shows the behaviour of the scalar field (along the vertical axis) and the second one shows the variation of the damping force (along the vertical) with respect to the extra dimensional coordinate (along the horizontal axis) when $V(\phi) = \frac{1}{2}m^{2}\phi^{2}$ and $f(y)= -A\ln\cosh(y)$.}
\label{Fig 2}
\end{figure}

For the decaying warping function viz. $f(y)= -A\ln\cosh(y)$, the behaviour of the scalar field and the damping force is illustrated in FIG.~\ref{Fig 2}.
Here the dissipative force decreases rapidly as we move towards the brane and therefore trapping will not take place.

\subsection{The potential $V(\phi) = \frac{1}{2}m^{2}\phi^{2}+\frac{\lambda}{4}\phi^{4}$}
\begin{figure}[ht]
\includegraphics[height=1.5in]{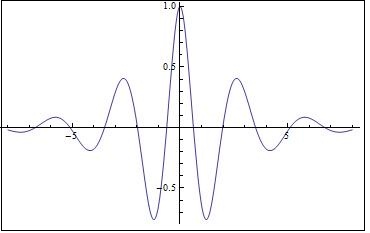}
\includegraphics[height=1.5in]{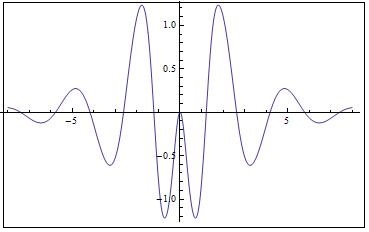}
\caption{Left figure describes the nature of the scalar field and right one describes the nature of the damping force with respect to the extra dimensional coordinate when $V(\phi) = \frac{1}{2}m^{2}\phi^{2} +\frac{\lambda}{4} \phi^{4}$ and $f(y)= A\ln\cosh(y)$.}
\label{Fig 3}
\end{figure}

Here we shall consider a single quadratically self coupled scalar field with $\lambda$ as the coupling constant. We may choose $\lambda$ to be either positive or negative. But if we choose a negative coupling constant, then it leads to a potential which has no lower bound. The potential has a single minimum at the origin if we assume that $m^{2}$ is positive for a positive coupling constant, and it will also remain invariant under $Z_{2}$ symmetry. But spontaneous symmetry breaking occurs when $m^{2}$ is negative \cite{spo1,spo2}. Thus we have considered both positive and negative values of $m^{2}$ in our analysis. The consideration of spontaneous symmetry breaking can lead to important consequences during the creation and formation of the universe. Symmetry breaking methods can be used to describe the production of particles without the loss of energy and gives rise to the gauge bosons. The field equation under such a potential can be written as
\begin{equation}\label{05a}
% \nonumber to remove numbering (before each equatio
\phi''(y)+ 4f'(y)\phi'(y) + m^{2}\phi(y)+\frac{\lambda}{3}\phi^{3}(y)= 0.
\end{equation}
For the growing warping function, FIG.~\ref{Fig 3} shows that both the field and the damping force are oscillatory in nature, but the oscillations die out gradually as we move away from the brane. The damping vanishes not only on the brane but also at other places along the extra dimension. Thus the system has several unstable points in addition to the one on the brane. In this case we do not expect any efficient confinement of particles on the brane.

 \begin{figure}[ht]
\includegraphics[height=1.5in]{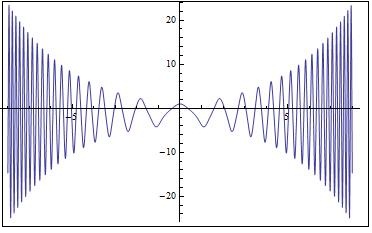}
\includegraphics[height=1.5in]{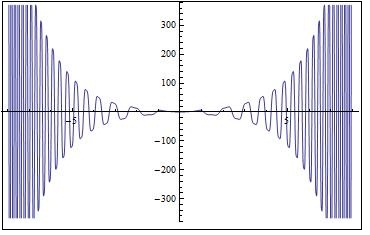}
\caption{First picture describes the nature of the scalar field and the second one describes the nature of the damping force along the extra dimension when $V(\phi) = \frac{1}{2}m^2\phi^{2} +\frac{\lambda}{4} \phi^{4}$ and $f(y)= -A\ln\cosh(y)$.}
\label{Fig 4}
\end{figure}

In the case of the decaying warping function, both the field and the damping shows oscillatory behaviour (FIG.~\ref{Fig 4}). But unlike the previous case, here the amplitude of oscillations diverge as we move away from the brane. The damping is absent on the brane.

\begin{figure}[ht]
\includegraphics[height=1.5in]{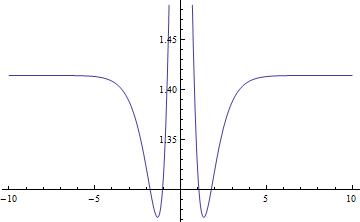}
\includegraphics[height=1.5in]{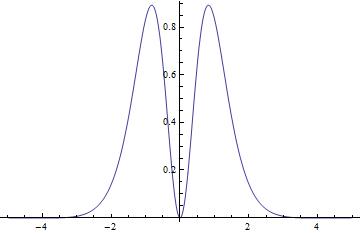}
\caption{The figure on the left describes the nature of the scalar field and the right one describes the nature of the damping force with respect to the extra dimensional coordinate when $V(\phi) = -\frac{1}{2}m^{2}\phi^{2} +\frac{\lambda}{4} \phi^{4}$ and $f(y)= A\ln\cosh(y)$.}
\label{Fig 3a}
\end{figure}

 \begin{figure}[ht]
\includegraphics[height=1.5in]{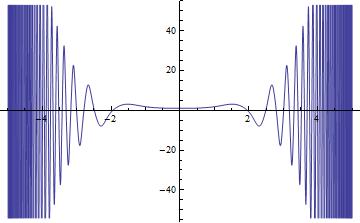}
\includegraphics[height=1.5in]{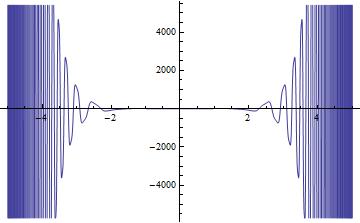}
\caption{First picture describes the nature of the scalar field and the second one describes the nature of the damping force along the extra dimension when $V(\phi) = -\frac{1}{2}m^2\phi^{2} +\frac{\lambda}{4} \phi^{4}$ and $f(y)= -A\ln\cosh(y)$.}
\label{Fig 4a}
\end{figure}

From FIG.~\ref{Fig 3a} it is clear that the presence of negative $m^{2}$ term implies that the dissipative nature of the system gradually falls as we move away from the brane in the case of the growing warping function. For the decaying type warping function, the dissipation is oscillating in nature along the extra dimensional coordinate.

\subsection{When $V(\phi) = C/\phi^{2}$}
\begin{figure}[ht]
\includegraphics[height=1.5in]{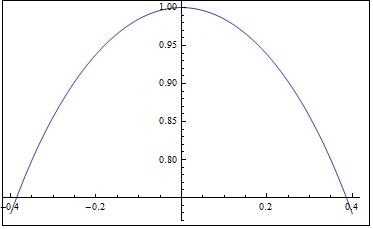}
\includegraphics[height=1.5in]{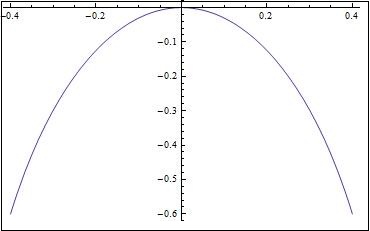}
\caption{The figure on the left describes the behaviour of the scalar field and the figure on the right describes the behaviour of the damping force with respect to the extra dimensional coordinate for $V(\phi) =C/\phi^{2}$ and $f(y)= A\ln\cosh(y)$.}
\label{Fig 5}
\end{figure}

\begin{figure}[ht]
\includegraphics[height=1.5in]{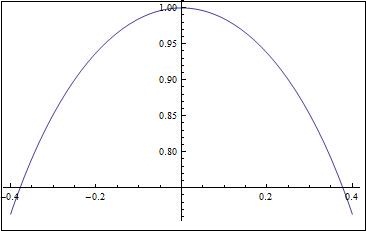}
\includegraphics[height=1.5in]{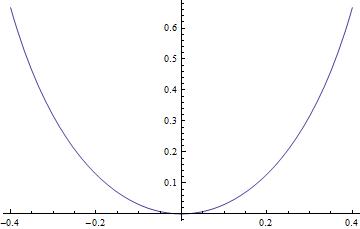}
\caption{First picture describes the nature of the scalar field and second one describe the nature of the damping force with respect to the extra dimensional coordinate when $V(\phi) = C/\phi^{2}$ and $f(y)= -A\ln\cosh(y)$.}
\label{Fig 6}
\end{figure}

Here the scalar field equation can be written as
 \begin{equation}\label{05b}
% \nonumber to remove numbering (before each equatio
\phi''(y)+ 4f'(y)\phi'(y) + C_{1}\phi^{-3}(y)= 0.
\end{equation}
FIG.~\ref{Fig 5} and FIG.~\ref{Fig 6} describe the scalar field and the damping for a growing and a decaying type warping function respectively. We can see that the variation of the scalar field is independent of whether the warping function is of growing or of decaying type, but the damping force plays opposite roles in the two cases. Although the system tends to stabilize for the growing warping function, but the damping is absent on the brane.

\subsection{Nature of the effective potential with interactive scalar field and the role of warping function}
\begin{figure}[ht]
\includegraphics[height=1.5in]{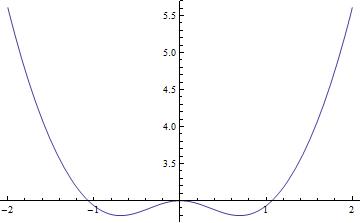}
\includegraphics[height=1.5in]{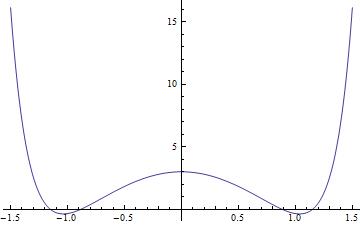}
\caption{First picture shows the variation of the effective potential (along the vertical) when $f(y)= A\ln\cosh(y)$ and second one shows the variation of the effective potential (along the vertical) when $f(y)= -A\ln\cosh(y)$ with respect to the extra dimensional coordinate (along the horizontal) for $V(\phi) = \frac{1}{2}m^2\phi^{2}$.}
\label{Fig 7}
\end{figure}

\begin{figure}[ht]
\includegraphics[height=1.5in]{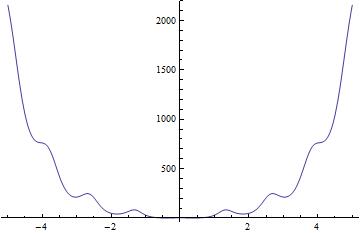}
\includegraphics[height=1.5in]{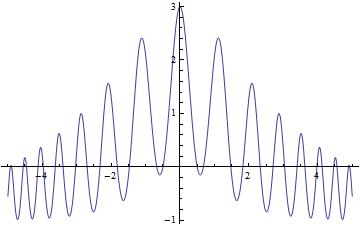}
\caption{First picture describes the variation of the effective potential when f(y)= A$\ln\cosh(y)$ and second one describe the variation of the effective potential when f(y)= -A$\ln\cosh(y)$  with respect to the extra dimensional coordinate for $V(\phi) = \frac{1}{2}m^{2}\phi^{2}+\frac{\lambda}{4}\phi^{4}$. }
\label{Fig 8}
\end{figure}

\begin{figure}[ht]
\includegraphics[height=1.5in]{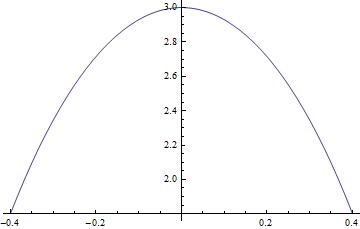}
\includegraphics[height=1.5in]{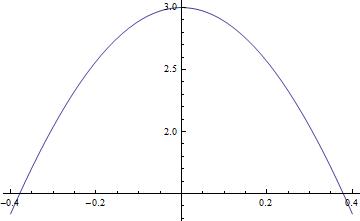}
\caption{First picture shows the variation of the effective potential when f(y)= A$\ln\cosh(y)$ and second one shows the variation of the effective potential when f(y)= -A$\ln\cosh(y)$ with respect to the extra dimensional coordinate for $V(\phi) = C/\phi^{2}$.}
\label{Fig 9}
\end{figure}

According to \cite{rs12a}, during a Yukawa type interaction between a test particle and a scalar field, the effective potential can be written as
\begin{equation}\label{05c}
% \nonumber to remove numbering (before each equatio
V_{eff}= \left[e^{2f(y)}\left(1+\frac{h^{2}\phi^{2}}{M^{2}}\right)-1\right],
\end{equation}
where $h$ is the coupling constant and $M$ is the mass of the test particle. As the metric used by us is identical to the one used in \cite{rs12a}, we have used the same expression of effective potential. However, the subsequent investigations undertaken here were not done in \cite{rs12a}. As we are unable to determine $\phi$ explicitly, once again we have used the NDsolve routine of Mathematica package to overcome the situation so as to compute the effective potential. The result of our exercise is presented in FIG.~\ref{Fig 7}, FIG.~\ref{Fig 8} and FIG.~\ref{Fig 9}. FIG.~\ref{Fig 7} shows that the effective potential is not minimum on the brane if we choose a growing or a decaying warping function, when $V(\phi)=\frac{1}{2}m^{2}\phi^{2}$. The effective potential is neutral on the brane for the growing warping function, but is repulsive for the decaying case. Hence trapping is impossible in the second case.

FIG.~\ref{Fig 8} shows that the effective potential decreases as we move towards the brane in the case of the growing warping function, although it fluctuates about specific positions along the extra dimension and becomes neutral on the brane. On the other hand a decaying warping function generates an oscillatory effective potential, but in such a case, as we move away from the brane, the magnitude of the effective potential will decrease. Although there will be no confinement on the brane, confinement may take place at specific regions along the extra dimension. In both these cases we have chosen $V(\phi)= \frac{1}{2}m^{2}\phi^{2}+\frac{\lambda}{4}\phi^{4}$.

FIG.~\ref{Fig 9} shows that for $V(\phi)= C/\phi^{2}$ the effective potential is indifferent to the nature of warping function, i.e. whether it is growing or decaying and is repulsive on the brane in both these cases.

\section{Stability of the dynamical system}
In this section we investigate the stability of a (i) warped system (ii) an unwarped system and (iii) a perturbed system. The study of the dynamical system under these three conditions helps us to understand the effect of warping and perturbation on the confinement of particles.
\subsection{System with warping }
\begin{figure}[ht]
\includegraphics[height=1.5in]{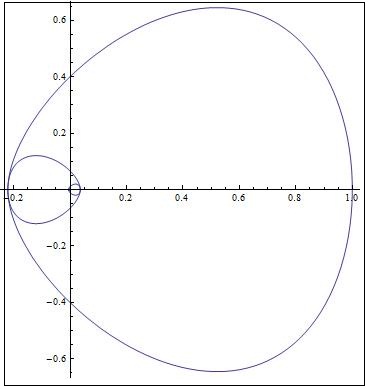}
\includegraphics[height=1.5in]{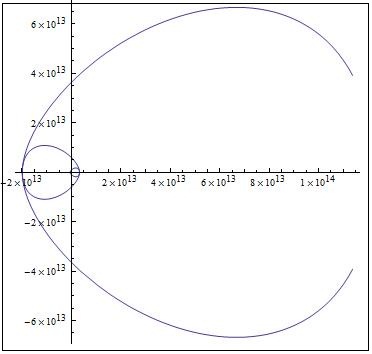}
\caption{First picture describes the nature of the periodic orbit when f(y)= A$\ln\cosh(y)$ and second one shows it for f(y)= -A$\ln\cosh(y)$ when $V(\phi) = \frac{1}{2}m^{2}\phi^{2}$.}
\label{Fig 10}
\end{figure}

In the previous sections we have studied the system under warped condition. FIG.~\ref{Fig 10} describes the system under growing and decaying warping functions. It is clear from the figure that the phase trajectories are not limit cycles, and hence the system is not in an asymptotically stable condition.

\subsection{System without warping}
In the absence of the warping function, the system will turn into a conservative one in behaviour \cite{to1a}. So it will not dissipate any energy \cite{rs18}, and as a result it may have some closed orbits but these orbits will not be isolated in nature. Under such a situation if we consider a scalar field perturbation (which also perturbs the scalar field potential), the system may produce a limit cycle, which is a self-sustained oscillation where energy dissipated over one cycle balances the energy gained during the cycle. Thus the presence of a limit cycle ensures an asymptotically stable dynamical system.

Here the unperturbed scalar field equation can be written as
\begin{equation}
% \nonumber to remove numbering (before each equatio
\phi''(y)+\frac{dV(\phi)}{d\phi}= 0.
\end{equation}
The corresponding dynamical system is given by
\begin{eqnarray}
  \phi' &=& q , \\
  q' &=& -\frac{dV(\phi)}{d\phi}.
\end{eqnarray}
It is evident that the behaviour of the system is largely dependent on the nature of potential. We shall perform the stability analysis for the system with quadratic potential and Higgs potential only. During the study of stability we have restricted ourselves to a Hamiltonian system. Hence we did not study the system under the potential $ V(\phi) = C/(\phi)^2$, which does not generate a Hamiltonian system.

\subsubsection{Unperturbed system with potential $V(\phi)=\frac{1}{2}m^{2}\phi^{2}$}
\begin{figure}[ht]
\includegraphics[height=2.0in]{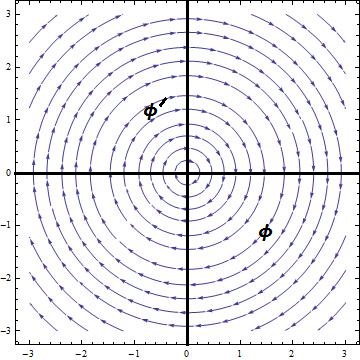}
\caption{Phase portrait for unperturbed and unwarped condition with $V(\phi)=\frac{1}{2}m^{2}\phi^{2}$.}
\label{Fig 11}
\end{figure}

The system in this situation is described by the set
\begin{eqnarray}
      \phi' &=& q , \\
      q' &=& -m^{2}\phi .
\end{eqnarray}
For simplicity if we assume $m^{2}=1$, then the system can be described by a one parameter family of periodic orbits $\gamma_{\alpha}(y) = (\alpha \cos(y), \alpha \sin(y))$ of period $2\pi$ (FIG.~\ref{Fig 11}). The solution curves obtained here are stable solutions, but not asymptotically stable.

\subsubsection{Perturbed system with quadratic potential}
It is known that perturbation can be generated during inflation \cite{rs16a,rs16b}. In most of these cases, the effect of the perturbation is judged by perturbing the metric. In our earlier work \cite{BG}, we considered the effect of metric perturbation on geodesic motion. But here we shall consider that the scalar field is perturbed due to a bending of the brane along the extra dimensional coordinate, which in its turn perturbs the corresponding field potential. Such a bending may occur due to the expansion of the brane before the stabilization of extra dimensions takes place, leading to a successful inflationary cosmology \cite{Kaloper} or due to a variable cosmological term which is large in the early universe but subsequently rolls down to attain a negligible magnitude at present \cite{MS}. In our study we shall assume that the perturbed scalar field can be expanded in the form of a Taylor's series about the brane i.e. about $y=y_0$ as follows:
\begin{equation}\label{11}
\phi(y+\epsilon) =  \phi(y)+ \left. \epsilon\frac{ d\phi(y)}{dy}\right|_{y_0} + \left. \frac{ \epsilon^{2}}{2}\frac{d^2 \phi(y)}{dy^2}\right|_{y_0} + \ldots ,
\end{equation}
or
\begin{equation}\label{11a}
\phi(y+\epsilon) =  \phi(y)+ \left. \epsilon q \right|_{y_0} + \left. \frac{ \epsilon^{2}q'}{2}\right|_{y_0} + \ldots ,
\end{equation}
where $\epsilon$ is a small quantity. The perturbed potential is found to be given by
\begin{eqnarray}
V(\phi)|_{p} &=& \frac{1}{2}m^{2}(\phi(y)+ \epsilon q)^{2},
\end{eqnarray}
where we have neglected the terms of second order of smallness. Therefore the dynamical system is now given by the following set:
\begin{eqnarray}
\phi' &=& q, \\
 q' &=& -\phi -\epsilon q.
\end{eqnarray}
We note that even if we retain the second order terms in the potential, no such term appears in the expression for $q'$.
We will now use the Melnikov theory to study the above dynamical system. It gives us an opportunity to study homoclinic loop bifurcations and establish the existence of transverse homoclinic orbits for perturbed dynamical systems. We can therefore check for the existence of limit cycles and separatrix cycle for the perturbed systems. To do so we have to calculate the Melnikov function. The Melnikov function is a measure of the separation of the stable and unstable manifolds at a point on the unperturbed homoclinic path.

\begin{figure}[ht]
\includegraphics[height=2.0in]{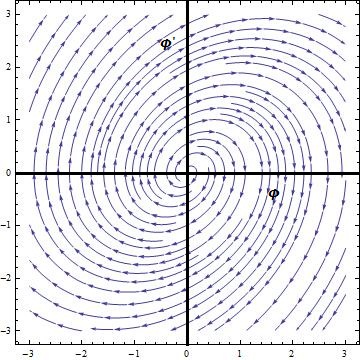}
\caption{Phase portrait for perturbed system with $V(\phi)=\frac{1}{2}m^{2}\phi^{2}$.}
\label{Fig 11a}
\end{figure}

The Melnikov function \cite{rs17,rs18} for the above case is obtained in the form
\begin{equation}
% \nonumber to remove numbering (before each equatio
M(\alpha)= \int^{2\pi}_{0}[\alpha^{2}\cos^{2}y]dy,
\end{equation}
or
\begin{equation}
% \nonumber to remove numbering (before each equatio
M(\alpha)= \alpha^{2} \pi ,
\end{equation}
which is therefore a finite positive quantity. So the the stable and unstable manifolds will not intersect and as a result there will be no limit cycle in this system. The corresponding phase portrait is given in (FIG.~\ref{Fig 11a}).

\subsection{System with Higgs potential}
The Higgs potential is expressed as $V(\phi) = -\mu_{1} \phi^{2}+ \mu_{2} \phi^{4}$. The plot of the variation of this potential energy with the scalar field is found to possess the shape of a Mexican hat \cite{quartic}. In particular, the minimum of energy value is not at $\phi$= 0. Instead it has an infinite number of possible minima (vacuum states) given by $\phi= \frac{\mu_{1}}{\sqrt{2\mu_{2}}}e^{i\theta}$ for any real $\theta$ between $0$ and 2$\pi$.

\subsubsection{Unperturbed system with Higgs potential}

\begin{figure}[ht]
\includegraphics[height=2.0in]{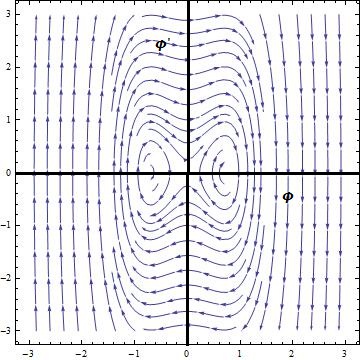}
\caption{Phase portrait for unperturbed system with Higgs potential.}
\label{Fig 11b}
\end{figure}

The system under such a potential is described by the set of equations
\begin{eqnarray}
\phi' &=& q , \\
 q' &=& 2\mu_{1}\phi -4\mu_{2}\phi^{3} .
\end{eqnarray}
This system clearly describes two homoclinic orbits given by $\gamma^{\pm}_{0}(y)= \pm (\frac{\sqrt{2}p}{\cosh y}, -\sqrt{2}p \frac{\tanh y}{\cosh y})$, with saddle at the origin and centers at $\pm p$, where $p = \sqrt{\frac{\mu_{1}}{\mu_{2}}}$. FIG.~\ref{Fig 11b} shows the phase portrait of the unperturbed system with the Higgs potential.

\subsubsection{Perturbed system with Higgs potential}
\begin{figure}[ht]
\includegraphics[height=2.0in]{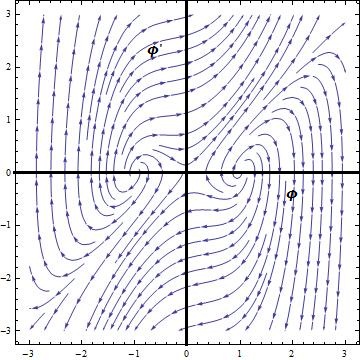}
\caption{Phase portrait for perturbed system with Higgs potential.}
\label{Fig 11c}
\end{figure}

If we introduce perturbation in a system with Higgs potential, the system assumes the form
\begin{eqnarray}
      \phi' &=& q , \\
      q' &=& 2\mu_{1}\phi -4\mu_{2}\phi^{3}- \epsilon[(2\mu_{1}q+ 12\mu_{2}\phi^{2}q + 6 \epsilon \mu_{2} q^{2} \phi].
\end{eqnarray}
We find that in this case the second order terms appear in the expression for $q'$. Thus the Melnikov function along the separatrix $\gamma^{+}_{0}(y)$ is obtained as
\begin{equation}\label{2}
  M(\mu,\epsilon)= \int^{\infty}_{-\infty}q[(2\mu_{1}q+ 12\mu_{2}\phi^{2}q + 6 \epsilon \mu_{2} q^{2} \phi]dy.
\end{equation}
This integral gives us $  M(\mu,\epsilon)= - \frac{8\mu_{1}p^{2}}{3} - \frac{64\mu_{2}p^{4}}{5} + 3\epsilon\mu_{2}p^{3}\pi\sqrt{2}$. Thus we can say that $ M(\mu)= 0$ if and only if $\mu_{1} = \mu_{2}\frac{33 \sqrt{2}\epsilon}{98}$. Therefore for all sufficiently small $\epsilon$, there is a $\mu_{\epsilon} = \mu_{2}(1,\frac{33 \sqrt{2} \epsilon}{98})$ such that the system has two homoclinic orbits $\gamma^{\pm}_{\epsilon}$ at the saddle point at the origin, in an $\epsilon$- neighbourhood of $\gamma^{\pm}_{0}$. The stability of these separatrix depends on the term $\epsilon \mu_{1} $. Since the separatrix cycles $\gamma^{\pm}_{\epsilon}\bigcup \{0\}$ are stable on their interior, thus a stable limit cycle is generated on the interior of these separatrix as $\mu_{2}$ decreases from $\frac{33\sqrt{2} \epsilon}{98}$ and a stable limit cycle on the exterior of the graphic $\gamma^{+}_{\epsilon}\bigcup \gamma^{-}_{\epsilon}\bigcup \{0\}$ as  $\mu_{1}$ increases from $\frac{33\sqrt{2} \epsilon}{98}$. The corresponding phase portrait is shown in FIG.~\ref{Fig 11c}.

\section{Summary}
Finally, we list below, systematically, the conclusions which we can draw from this piece of investigation.

$\bullet$ In this paper we have studied the behavior of a scalar field in a 5D warped product spacetime. The warping function is dependent on the extra-dimensional coordinate only. Similarly, the scalar field we have chosen here, is a function of the extra-dimensional coordinate. The quantity $V(\phi)$ represents the potential due to the scalar field. The behaviour of the dynamical system has been studied for different scalar field potentials in order to get some idea about the nature of confinement.

$\bullet$  The whole study have been divided into three parts. In the first part we have analyzed the behaviour of the scalar field and the associated damping force. The damping force is an additional force which helps the system to attain equilibrium. In our study, the damping force depends on the extra dimensional coordinate. First we considered the dynamical system in presence of a massive potential $V(\phi) = \frac{1}{2}m^2 \phi^{2} $. Here the damping force becomes maximum at the brane for a growing type of warping function, but the maximum value is zero. This indicates that if we move away from the the brane the damping will de-stabilize the system. For a decaying warping function the effect is just the opposite. Next we studied the system under the action of a quartic potential $V(\phi) = \frac{1}{2}m^2 \phi^{2} +\frac{\lambda}{4} \phi^{4}$. Here the the system finds itself in stable, neutral and unstable conditions periodically for both growing and decaying warp factors as we move along the extra dimension. A growing warping function helps the oscillations of the system to die out gradually as we move away from the brane, whereas for a decaying warping function, the oscillations diverge as we move away from the brane. However, for both types of warping function, the damping force vanishes on the brane. Lastly we have studied the effect of the potential $V(\phi) = C/\phi^{2} $. Here the nature of the scalar field remains almost the same for both growing and decaying warping functions, as if the field is indifferent to the warp factor, though the corresponding damping force shows opposite behaviour in these two cases. Although the system tends to stabilize for the growing warping function, but the damping is absent on the brane.

$\bullet$ In the second part we have assumed the scalar field to have Yukawa-type interaction with the test particle. Under this condition we have studied the effective potential, which shows a double bottom nature for growing and decaying warping functions when the scalar field has a massive quadratic potential. For a quartic potential, in the case of a growing warping function, the effective potential becomes minimum at the brane, but the minimum value is zero, indicating neutral confinement. Whereas for the decaying warping function it attains a repulsive maximum on the brane. But the interesting part appears when we choose $V(\phi) = C/\phi^{2} $. The behaviour of the effective potential on the brane is identical for both the warping functions, and is repulsive in nature. This shows that the system ignores the presence of growing and decaying warping functions in the presence of this scalar field potential.

$\bullet$ In the final part we have studied the stability of the dynamical system in presence and absence of a perturbation during an unwarped condition. We found that the presence of perturbation may or may not generate asymptotically stable cycle i.e a limit cycle, depending on the nature of the scalar field potential. In the warped condition, the phase trajectories are not limit cycles and therefore we need not analyze the stability of the system in this condition. In the unwarped case, for the quadratic potential we find that there is a condition on the possibility of obtaining a limit cycle when the system remains unperturbed. So we conclude that there is a restriction on the situation when the system can become asymptotically stable. For the perturbed system the stable and unstable manifolds does not intersect and as a result there will be no limit cycle. During the study of the Higgs potential we found that we may obtain stable and unstable limit cycles.

Summarising we can say that the above analysis throws new light on the issue of confinement in presence of scalar fields in the bulk 5D spacetime for different types of scalar field potentials. The nature of the dissipation has been studied to see the role of dissipative term in confinement. During the study of the conservative system we have considered the effect of a perturbation to see whether, and under what condition dissipation comes into play in the system. To do so we determined the condition to get a limit cycle which indicates an asymptotically stable dynamical system.

\section*{Acknowledgments}
SG thanks IUCAA, India for an associateship. PB gratefully acknowledges the facilities available at the Department of Physics, St. Xavier's College (Autonomous), Kolkata and the Relativity and Cosmology Research Center, Jadavpur University.

\end{document}